
%
\input phyzzx
\tolerance=1000
\sequentialequations
\def\rl{\rightline}

\def\r#1{$\bf#1$}

\def\t1{{\tilde 1}}

\def\AEF{A.E. Faraggi}
\def\DVN{D. V. Nanopoulos}

\def\SUSY{supersymmetry }

\def\NPB#1#2#3{Nucl. Phys. B {\bf#1} (19#2) #3}
\def\PLB#1#2#3{Phys. Lett. B {\bf#1} (19#2) #3}
\def\PRD#1#2#3{Phys. Rev. D {\bf#1} (19#2) #3}

\def\l{\langle}
\def\r{\rangle}

\REF\GSW{M. Green, J. Schwarz and E. Witten,
Superstring Theory, 2 vols., Cambridge
University Press, 1987.}
\REF\SUGRA{H. P. Nilles, Phys. Rep. {\bf 110} (1984) 1; D. V. Nanopoulos and
A. B. Lahanas, Phys. Rep. {\bf 145} (1987) 1.}
\REF\WIT{E. Witten, \PLB{155}{85}{151}.}
\REF\CFT{R. Dijkgraaf, E. Verlinde and H. Verlinde, Comm. Math. Phys. {\bf
115} (1988) 649; L. Dixon, P. Ginsparg and J. Harvey, \NPB{306}{88}{470}.}
\REF\MOD{K. Narain, M. Sarmadi and E. Witten, \NPB{279}{87}{369}.}
\REF\MO{S. Cecotti, S. Ferrara and L. Girardello, \NPB{308}{88}{436}.}
\REF\TSD{A. Giveon, M. Poratti and E. Rabanovici, Phys. Rep. {\bf 244} (1994)
77.}
\REF\SUSY{J. P. Deredinger, L. E. Ibanez and H. P. Nilles, \PLB{155}{85}{65};
M. Dine, R. Rhom, N. Seiberg and E. Witten, \PLB{156}{85}{55}.}
\REF\CYM{M. Dine and N. Seiberg, \NPB{301}{88}{357}; N. Seiberg, \NPB{303}
{88}{286}.}
\REF\ORB{L. Dixon, J. Harvey, C Vafa and E. Witten, \NPB{261}{86}{678};
\NPB{274}{86}{285}.}
\REF\ORBM{M. Cvetic, J. Louis and B. A. Ovrut, \PLB{206}{88}{227}; M. Cvetic,
J. Molera and B. A. Ovrut, \PRD{40}{89}{1140}.}
\REF\FFFM{I. Antoniadis, J. Ellis, E. Floratos, D. V. Nanopoulos and T.
Tomaras, \PLB{191}{87}{96}; S. Ferrara, L.Girardello, C. Kounnas and M.
Poratti, \PLB{194}{86}{263}.}
\REF\NLY{J. L. Lopez, D. V. Nanopoulos and K. Yuan, CERN-TH.7259/94,
CTP-TAMU-14/94,hep-th/9405120.}
\REF\AFI{J. Bagger, N. Nemeschansky, N. Seiberg and S. Yankielowicz, \NPB{289}
{87}{53}.}
\REF\SLM{\AEF, \NPB{387}{92}{289}; \PLB{278}{92}{131}.}
\REF\FFF{I. Antoniadis, C. Bachas, and C. Kounnas, \NPB{289}{87}{87};
I. Antoniadis and C. Bachas, \NPB{298}{88}{586};
H. Kawai, D.C. Lewellen, and S.H.-H. Tye,
Phys. Rev. Lett. {\bf57} (1986) 1832;
Phys. Rev. D {\bf 34} (1986) 3794;
Nucl. Phys. B {\bf 288} (1987) 1;
R. Bluhm, L. Dolan, and P. Goddard,
Nucl. Phys. B {\bf 309} (1988) 330.}
\REF\KLN{S. Kalara, J. Lopez and D.V. Nanopoulos,\PLB{245}{91}{421};
\NPB{353}{91}{650}.}
\REF\REVAMP{I. Antoniadis, J. Ellis, J. Hagelin, and \DVN, \PLB{231}{89}{65}.}
\REF\NAR{K. Narain, \PLB{169}{86}{41}.}
\REF\ZT{\AEF, \PLB{326}{94}{62}.}
\REF\IL{L. E. Ibanez and D. Lust, \NPB{382}{92}{305}.}
\REF\ASY{K. Narain, M. Sarmadi and C. Vafa, \NPB{288}{87}{551}.}

\singlespace
\rl{WIS--94/41/SEP--PH}
\rl{\today}
\rl{T}
\normalspace
\smallskip
\titlestyle{\bf{Untwisted Moduli and Internal Fermions in
Free Fermionic Strings}}
\author{Edi Halyo
{\footnote*{e--mail address: jphalyo@weizmann.bitnet.}}}
\smallskip
\centerline {Department of Particle Physics}
\centerline {Weizmann Institute of Science}
\centerline {Rehovot 76100, Israel}
\bigskip
\titlestyle{\bf ABSTRACT}

We investigate the dependence of the number and type of untwisted moduli
on the boundary condition vectors of relistic free fermionic strings. The
number of moduli is given by six
minus the number of complex internal world--sheet fermions and the type of
moduli is determined by the details of the world--sheet left--right
asymmetry of the boundary conditions
for the internal fermions. We give a geometrical description of our results
in terms of the
transformations of the compactified dimensions of $Z_2 \times
Z_2$ orbifolds. We investigate all possible boundary conditions for the
internal fermions and
prove our results in general by showing that world--sheet
supersymmetry eliminates those boundary conditions which violate our results.

\singlespace
\vskip 0.5cm
\nopagenumbers
\pageno=0
\endpage
\normalspace
\pagenumbers

\centerline{\bf 1. Introduction}

The low--energy limit of superstrings [\GSW] are given by $N=1$ supergravity
theories [\SUGRA] with a gauge group and particle content fixed by the
underlying string.
These supergravity theories are completely defined by three functions: the
gauge function $f_{\alpha \beta}$, the Kahler potential $K$, and the
superpotential $W$ which can be derived from the string theory (at least in
principle). The massless spectrum of superstrings
always contains particles called moduli. These are gauge singlet scalars
which do not have a potential to any order in perturbation theory. They
obtain a potential and mass only at the nonperturbative level. A modulus
always present in string models is the dilaton, $S$, with a Kahler potential
$K(S,{\bar S})=-log(S+{\bar S})$ which spans the coset space $SU(1,1)/U(1)$
[\WIT].
In general, there are other moduli fields whose number and properites
(such as Kahler potential and coset space) are model dependent. These are
divided into twisted and untwisted moduli depending on whether they arise
from the twisted (Ramond) or untwisted (Neveu--Shwarz) sectors of the massless
string spectrum. In this paper,
we investigate the untwisted moduli in four dimensional free fermionic
superstrings. Since the
properties of the dilaton are universal and well--understood, we will
concentrate on the other untwisted moduli.

A given string model is perturbed into another one when the moduli obtain VEVs.
In other words, the moduli parametrize a continuous family of string models
through their VEVs. On the other hand, every string model is defined by a
conformal field theory on the world--sheet[\GSW]. This conformal field theory
can be perturbed by exactly marginal operators to give a new one which
correponds to a different string model[\CFT]. Each untwisted modulus field
appears as the
coupling constant of an exactly marginal operator in the world--sheet string
action[\MOD,\MO]. Thus there is
a one--to--one correspondence between the untwisted moduli in space--time
and the exactly marginal operators on the world--sheet.

The study of untwisted moduli (other than the dilaton) is important both for
theoretical and phenomenological reasons. On the theoretical side, the coset
space of these moduli gives the family of string models which are continuously
connected to each other through the moduli VEVs. In addition, symmetries of
the string which are believed to be exact, such as target space duality[\TSD]
depend on the number and type of untwisted moduli present.
On the phenomenological side, the untwisted
moduli play an important role in low--energy physics such as supersymmetry
breaking. One of the most popular supersymmetry breaking scenarios is
due to nonzero moduli F terms arising from hidden sector gaugino condensation
[\SUSY]. This mechanism depends crucially on the Kahler potential of the
moduli which must also be obtained from the underlying string theory.
Therefore, it is important to know the number and type of moduli together with
their Kahler potential and the coset space they span.

For (2,2) Calabi--Yau models [\GSW] the untwisted moduli can be found by
a standard procedure[\CYM]. The moduli of (2,2) and (2,0) orbifold models
[\ORB] have been
investigated in Refs. [\ORBM]. It was found that the untwisted moduli of
orbifold models correspond to products of the Cartan currents of the
Kac--Moody algebra which
are compatible with the orbifold twist. In terms of the background fields,
the moduli are those fields which are preserve the point group of the orbifold.

Untwisted moduli of free fermionic strings were first investigated in Refs.
[\FFFM]
by using a truncation procedure similar to dimensional reduction. It was found
that, in addition to the dilaton, the Neveu--Shwarz sector of the massless
string spectrum contains other gauge singlet scalars which parametrize a coset
space. More recently the same issues were explored in Ref. [\NLY] on which we
rely strongly in this paper. It was shown that, in free fermionic models,
the untwisted moduli appear as the couplings of Abelian Thirring interactions
of the form $J^i_L {\bar J}_R^j$ on the world--sheet. These interactions which
perturb the original free fermionic string were shown to be exactly marginal
operators[\AFI].
The untwisted moduli of a given string model correspond to the Abelian Thirring
interactions which are compatible with the basis vectors or the spin structure
which define the model. As a result, the basis vectors fix the moduli of the
string model. The purpose of this work is to find how the basis vectors and the
boundary conditions which they define in a given free fermionic string model
fix the number and type of untwisted moduli present. Our results and
conclusions are
completely general even though for conceteness the examples we give in
the following are from standard--like superstring models[\SLM].

The paper is organized as follows. In Section 2, we briefly introduce the
four dimensional free fermionic formulation of heterotic strings. In Section
3, we review some of the results of Ref. [\NLY] which are essential in the
following. We show how Abelian Thirring interactions are used to identify the
untwisted moduli. We define the different types of moduli possible, their
Kahler potentials and the coset spaces they span. In Section 4, we show how
to translate the free fermionic formulation to the $Z_2 \times Z_2$ orbifold
formulation in order to get a geometrical picture. In Section 5, we consider
three examples of realistic superstring models with basis vectors
which correspond to asymmetric
Wilson lines. We show how the number and type of moduli depend on the basis
vectors or the spin structure in a given model. In Section 6, we give a
geometrical interpretation of the results of Section 5 by translating them to
the orbifold formulation. In Section 7, we generalize the results of Section 5
by considering all possible boundary conditions for the internal fermions.
We find all the possibilities for the remaining moduli when there are complex
internal fermions. We prove our results in general (for models with $Z_2$
twists) by showing that the boundary conditions which
violate them are not allowed by world--sheet supersymmetry. In Section
8, we give our conclusions.

\bigskip
\centerline{\bf 2. Superstrings in the free fermionic formulation}

In the four dimensional free fermionic formulation of superstrings[\FFF], all
world--sheet degrees of freedom required to cancel the conformal anomaly
are represented by free fermions on the world--sheet. The left--handed degrees
of freedom are the space--time fields $X^{\mu}$ ($\mu=0,1$ in the light--cone
gauge), their
supersymmetric partners $\psi^{\mu}$ and 18 real free fermions $\chi^i,y^i,
\omega^i$ ($i=1,\ldots,6$). The world--sheet supercurrent (in the
left--moving sector) is given by[\FFF]
$$T_F(z)=\psi^{\mu} \partial_z X_{\mu}+ \sum_{i=1}^6 {\chi^i y^i \omega^i}.
\eqno(2.1)$$
The right--handed world--sheet fields are ${\bar X}^{\mu}$ and 44 real
world--sheet fermions ${\bar \phi}^a$, $a=1,\ldots, 44$. Under parallel
transport around a noncontractible loop of the world--sheet, the fermions
pick up a phase (which make up the spin structure). A model in the free
fermionic formulation is defined by a set of basis vectors, $b_i$, of
boundary condition vectors for each world--sheet fermion and a set of GSO
coefficients
$c\left(\matrix{b_i\cr  b_j\cr}\right)$[\FFF]. The basis vectors and the
coefficients are constrained by modular invariance and have to satisfy a
set of conditions[\FFF]. If two world--sheet fermions (of the
same chirality) have the same boundary conditions in all basis vectors, then
they can be combined into a complex world--sheet fermion. On the other hand,
a right--handed and a left--handed world--sheet fermion with the same boundary
conditions can be combined to give a sigma model operator on the world--sheet.
It is important to note that sigma model operators correspond to left--right
symmetric boundary conditions whereas a complex fermion is equivalent to
asymmetric boundary conditions for the two real fermions which make it up.

A state $|s\r_{\alpha}$ is in the (sector $\alpha$ of the) physical Hilbert
space if it satisfies the generalized GSO projections[\FFF] for each basis
vector $b_i$
$$\left\{e^{i \pi (b_i F_{\alpha})}-\delta_{\alpha}c^*\left(\matrix{\alpha\cr
    b_i\cr}\right)\right\}|s\r_{\alpha}=0, \eqno(2.2)$$
where $\delta_{\alpha}=e^{i \alpha(\psi^{\mu})}$ and
$$(b_i F_{\alpha})=\{\sum_{left}-\sum_{right}\}(b_i(f) F_{\alpha}(f)).
\eqno(2.3)$$
Here, $F_{\alpha}(f)$ is a fermion number operator which counts each complex
fermion
$f$ once and $f^*$ minus once. For periodic world--sheet fermions there are two
degenerate states $|+\r,|-\r$ in the Ramond vacuum with $F=0,-1$ respectively.
For each complex fermion there is a space--time $U(1)$ symmetry corresponding
to the world--sheet current $f f^*$ with the charges
$$Q(f)={1\over 2}\alpha(f)+F(f), \eqno(2.4)$$
where $\alpha(f)$ is the boundary condition of fermion $f$ in the vector
$\alpha$. The massless spectrum of the string model is given by the states
which satisfy[\FFF]
$$M_L^2=-{1\over 2}+{{\alpha_L \cdot \alpha_L} \over 8}+N_L=-1+{{\alpha_R
\cdot \alpha_R} \over 8}+N_R=M_R^2=0, \eqno(2.5)$$
where $\alpha=(\alpha_L;\alpha_R)$ is a sector in the Hilbert space,
$N_L=\sum_f(\nu_L)$, $N_R=\sum_f(\nu_R)$ and $\nu_{f,f^*}=(1 \pm\alpha(f))/2$.
The trilevel and higher order terms in the superpotential, $W_n$, are obtained
from the world--sheet correlators
$$A_n=\l V_1^f V_2^f V_3^b \ldots V_n^b \r, \eqno(2.6)$$
by using the rules of Ref. [\KLN]. The nonvanishing correlators satisfy
all the world--sheet
selection rules due to the local and global charges and sigma models after
the necessary picture changings are done.

\bigskip
\centerline{\bf 3. Untwisted moduli in free fermionic models}

The problem of identifying the untwisted moduli in free fermionic strings has
been recently investigated. In this section, we review the results of Ref.
[\NLY].
This will enable us to establish the logic and method which will be essential
in the following
sections. In conformal field theory, moduli fields correspond to the exactly
marginal operators which deform the conformal field theory in a way which
preserves conformal invariance at the quantum level. In symmetric orbifolds
[\ORB],
for example, these exactly marginal operators which correspond to the untwisted
moduli are given by $\partial X^I \bar \partial X^J$ ($I=1, \ldots,6$, $J=1,
\ldots ,22$ and $X^I$ are the coordinates of the six torus $T^6$). In this
context, the untwisted moduli are the background fields [\MOD](the metric,
antisymmetric tensor and the Wilson lines) in the orbifold
construction which are compatible with the point group symmetry of the
orbifold.
The operators $i \partial X^I$ are the $U(1)$ Cartan generators of the
Kac--Moody algebra.

Analogously, one expects the exactly marginal opeartors in the free fermionic
models to correspond to Abelian Thirring operators $J^i_L(z)
{\bar J}^j_R(\bar z)$, $i=1, \ldots,6$, $j=1,
\ldots ,22$ . It has been shown that Abelian Thirring interactions
preserve conformal invariance and are therefore exactly marginal operators
[\AFI]. Thus, one can use them to identify the untwisted moduli in free
fermionic strings. The untwisted moduli correspond to the
Abelian Thirring interactions which are compatible with the spin structure
of the world--sheet fermions given by the boundary conditon vectors which
define the string model.

Consider first a basis ${\cal B}=\{{\bf 1},S\}$ where in vector ${\bf 1}$ all
world--sheet fermions are periodic and $S$ is the supersymmetry generator given
by
$$S=({\underbrace{1,\cdots,1}_{{\psi^\mu},{\chi^{1,...,6}}}},0,\cdots,0
\vert 0,\cdots,0). \eqno(3.1)$$
The two dimensional action for the Abelian Thirring interactions is[\NLY]
$$S^{\prime}=\int d^2z h_{ij}(X) J^i_L(z) {\bar J}^j_R(\bar z), \eqno(3.2)$$
where $i=1,\ldots,6$ and $j=1,\ldots,22$, corresponding to the left--handed
$U(1)^6$ and right--handed $U(1)^{22}$ Cartan subalgebras of the Kac--Moody
algebra. The $6 \times 22$ couplings $h_{ij}$ are the untwisted moduli fields
of the model which parametrize the coset space $SO(6,22)/SO(6) \times SO(22)$.
The moduli appear in the Neveu--Shwarz sector of the massless spectrum which
contains the following gauge singlet states (in addition to the gravity
multiplet and gauge bosons):
$$\eqalignno{&\chi^i|0\r \otimes {\bar \psi}^a {\bar \psi}^{-a}|0\r, &(3.3a)\cr
             &\chi^i|0\r \otimes {\bar \psi}^{\pm a} {\bar \psi}^{\pm b}|0\r,
&(3.3b)}$$
given in terms of the 22 complex right--handed world--sheet fermions
${\bar \psi}^a$
and their conjugates ${\bar \psi}^{-a}$, $a=1,\ldots,22$. The untwisted moduli
are given by the states in Eq. (3.3a) and correspond to the Abelian Thirring
interactions[\NLY]
$$J^i_L(z) {\bar J}^j_R(\bar z)=:y^i(z) \omega^i(z)::{\bar \psi}^a
{\bar \psi}^{-a}:, \eqno(3.4)$$
where $j=a$. Note that the boundary conditions of $\chi^i$ which appears in
the moduli and $y^i \omega^i$ which appears in the Abelian Thirring
interactions
are the same for a basis vector with $b_i(\psi^{\mu})=1$ due to the
tranformation properties of the supercurrent $T_F$. (If $b_i(\psi^{\mu})=0$,
then they are opposite.)
Next consider a model with the basis ${\cal B}^{\prime}=\{
{\bf 1},S,b_1,b_2,b_3\}$ where
$$\eqalignno{S&=({\underbrace{1,\cdots,1}_{{\psi^\mu},
{\chi^{1,...,6}}}},0,\cdots,0
\vert 0,\cdots,0),&(3.5a)\cr
b_1&=({\underbrace{1,\cdots\cdots\cdots,1}_
{{\psi^\mu},{\chi^{12}},y^{3,...,6},{\bar y}^{3,...,6}}},0,\cdots,0\vert
{\underbrace{1,\cdots,1}_{{\bar\psi}^{1,...,5},
{\bar\eta}^1}},0,\cdots,0),&(3.5b)\cr
b_2&=({\underbrace{1,\cdots\cdots\cdots\cdots\cdots,1}_
{{\psi^\mu},{\chi^{34}},{y^{1,2}},
{\omega^{5,6}},{{\bar y}^{1,2}}{{\bar\omega}^{5,6}}}}
,0,\cdots,0\vert
{\underbrace{1,\cdots,1}_{{{\bar\psi}^{1,...,5}},{\bar\eta}^2}}
,0,\cdots,0),&(3.5c)\cr
b_3&=({\underbrace{1,\cdots\cdots\cdots\cdots\cdots,1}_
{{\psi^\mu},{\chi^{56}},{\omega^{1,\cdots,4}},
{{\bar\omega}^{1,\cdots,4}}}},0,\cdots,0
\vert {\underbrace{1,\cdots,1}_{{\bar\psi}^{1,...,5},
{\bar\eta}^3}},0,\cdots,0),&(3.5d)\cr}$$
with the choice of generalized GSO projections
$$c\left(\matrix{b_i\cr
                                    b_j\cr}\right)=
c\left(\matrix{b_i\cr
                                    S\cr}\right)=
c\left(\matrix{1\cr
                                    1\cr}\right)=-1,\eqno(3.6)$$
and the others given by modular invariance.
This set of basis vectors is common to all realistic string models in the free
fermionic formulation such as the flipped $SU(5) \times U(1)$[\REVAMP] and
standard--like models[\SLM]. In the above notation the right--handed fermions
are
separated into 12 real ones ${\bar y}^i,{\bar \omega}^i$, $i=1,\ldots,6$ and
16 complex ones $\{{\bar \psi}^{1,\ldots,5},{\bar \eta}^{1,2,3},{\bar \phi}
^{1,\ldots,8}\}$. Thus the untwisted moduli fields given by Eq. (3.3a) are
separated into two sets
$$\eqalignno{&\chi^i|0\r \otimes {\bar y}^i {\bar \omega}^i|0\r, &(3.7a)\cr
             &\chi^i|0\r \otimes {\bar \psi}^a {\bar \psi}^{-a}|0\r, &(3.7b)}$$
which give the Cartan subalgebra of $SO(12) \times SO(32)$. Accordingly,
the Abelian Thirring interactions in Eq. (3.4) are also divided into two sets
$$\eqalignno{&J^i_L(z) {\bar J}^i_R(\bar z) \qquad i=1,\ldots,6, &(3.8a)\cr
             &J^i_L(z) {\bar J}^j_R(\bar z) \qquad j=6,\ldots,22. &(3.8b)}$$
The effect of
the new basis vectors $\{b_1,b_2,b_3\}$ is to make some of the world--sheet
currents $J^i_L$ or ${\bar J}^j_R$ antiperiodic. Then, some of the Abelian
Thirring interactions become not invariant under parallel transport of fermions
around noncontractible closed loops and are therefore forbidden. Thus, the
additional basis vectors eliminate the Thirring interactions which are not
invariant under their boundary conditions and only the ones which are
invariant remain as exactly marginal operators. As expected, the same basis
vectors also eliminate the corresponding moduli from the massless
string spectrum by the generalized GSO projections given by Eq. (2.2).

{}From the basis vector $b_1$ we find the boundary conditions
$$\eqalignno{&J_L^{1,2} \to J_L^{1,2}, \qquad J_L^{3,4,5,6} \to -J_L^{3,4,5,6},
&(3.9a)\cr
 &{\bar J}_R^{1,2} \to {\bar J}_R^{1,2}, \qquad {\bar J}_R^{3,4,5,6} \to -
{\bar J}_R^{3,4,5,6}, &(3.9b)}$$
and ${\bar J}^j_R$ ($j=7,\ldots,22$) periodic. The Thirring interactions
which remain invariant are
$$J_L^{1,2} {\bar J}_R^{1,2}, \qquad J_L^{3,4,5,6} {\bar J}_R^{3,4,5,6},
\qquad J_L^{1,2} {\bar J}_R^{7,\ldots,22}. \eqno(3.10)$$
Similarly under $b_2$ we obtain
$$\eqalignno{&J_L^{3,4} \to J_L^{3,4}, \qquad J_L^{1,2,5,6} \to -J_L^{1,2,5,6},
&(3.11a)\cr
 &{\bar J}_R^{3,4} \to {\bar J}_R^{3,4}, \qquad {\bar J}_R^{1,2,5,6} \to -
{\bar J}_R^{1,2,5,6}, &(3.11b)}$$
and ${\bar J}^j_R$ ($j=7,\ldots,22$) periodic. We are left with the Thirring
interactions (symmetric under both $b_1$ and $b_2$)
$$J_L^{1,2} {\bar J}_R^{1,2}, \qquad J_L^{3,4} {\bar J}_R^{3,4},
\qquad J_L^{5,6} {\bar J}_R^{5,6}. \eqno(3.12)$$
Boundary conditions under $b_3$ do not eliminate any of the terms in Eq. (3.12)
and therefore the Abelian Thirring interactions allowed by the basis vectors
$\{b_1,b_2,b_3\}$ are those in Eq. (3.12). In  terms of the massless string
states given in Eq. (3.3a), the moduli which correspond to the above marginal
operators are
$$h_{ij}=\chi^i|0\r \otimes {\bar y}^j {\bar \omega}^j |0\r \left \lbrace
{\matrix{(i,j=1,2)\cr (i,j=3,4)\cr (i,j=5,6)} \qquad \qquad \qquad \qquad
\qquad \eqno(3.13)$$
{}From the 132 real untwisted moduli given in Eq. (3.7) only these 12 real ones
remain after the basis vectors $\{b_1,b_2,b_3\}$ are taken into account.
These moduli are found among the completely neutral states of the
Neveu--Shwarz sector[\SLM]
$$\eqalignno{&\chi^{12}|0\r \otimes \{{\bar \omega}^{1,2,3,4} \}
\{ {\bar y}^{1,2} {\bar \omega}^{5,6}\} |0\r, &(3.14a)\cr
&\chi^{34}|0\r \otimes \{{\bar y}^{3,4,5,6} \} \{ {\bar \omega}^{1,2,3,4}\}
|0\r, &(3.14b)\cr
&\chi^{56}|0\r \otimes  \{{\bar y}^{3,4,5,6} \} \{ {\bar y}^{1,2} {\bar
\omega}^{5,6}\} |0\r, &(3.14c)}$$
after the GSO projections due to $\{b_1,b_2,b_3\}$ are
applied. Here $\chi^{ij}=\chi^i+i\chi^j$ and $|0\r$ is the Neveu--Shwarz
vacuum. The moduli in Eq. (3.13) span the coset (moduli) space[\NLY]
$${\cal M}=\left({SO(2,2) \over {SO(2) \times SO(2)}} \right)^3. \eqno(3.15)$$
We can form 6 complex moduli from the 12 real ones by
$$\eqalignno{&H_1^{(1)}={1\over \sqrt2}(h_{11}+ih_{21})={1\over \sqrt2}
(\chi^1+i \chi^2) |0\r \otimes {\bar y}^1 {\bar \omega}^1 |0\r, &(3.16a) \cr
       &H_2^{(1)}={1\over \sqrt2}(h_{12}+ih_{22})={1\over \sqrt2}
(\chi^1+i \chi^2)|0\r \otimes {\bar y}^2 {\bar \omega}^2 |0\r, &(3.16b)}$$
for the first set and analogously for the other two sets $H_{1,2}^{(2)},
H_{1,2}^{(3)}$. The Kahler potential for $H_j^{(i)}$ is[\NLY]
$$K(H,{\bar H})=-\sum_{i=1}^3 log \left(1-\sum_{j=1,2} H_j^{(i)} {\bar H}
_j^{(i)}+{1\over 4} |\sum_{j=1,2} H_j^{(i)}  H_j^{(i)}|^2 \right).
\eqno(3.17)$$
These in turn are connected to the moduli which give the Kahler structure
($T_i$) and the complex structure ($U_i$) by
$$\eqalignno{&T_1={1\over \sqrt2}(H_1^{(1)}-iH_2^{(1)})={1\over \sqrt2}
(\chi^1+i \chi^2) |0\r \otimes ({\bar y}^1 {\bar \omega}^1-i
{\bar y}^2 {\bar \omega}^2) |0\r, &(3.18a) \cr
       &U_1={1\over \sqrt2}(H_1^{(1)}+iH_2^{(1)})={1\over \sqrt2}
(\chi^1+i \chi^2) |0\r \otimes ({\bar y}^1 {\bar \omega}^1 +i
{\bar y}^2 {\bar \omega}^2) |0\r, &(3.18b)}$$
and analogously for $T_{2,3}$ and $U_{2,3}$.
The Kahler potential for $T_i$ and $U_i$ is given by
$$K(T_i, {\bar T}_i,U_i, {\bar U}_i)=-\sum_{i=1}^3 log (1-T_i {\bar T}_i)
-\sum_{i=1}^3 log (1-U_i {\bar U}_i), \eqno(3.19)$$
and they span the coset (moduli) space[\NLY]
$${\cal M}=\left({SU(1,1) \over U(1)} \otimes {SU(1,1) \over U(1)} \right)^3.
\eqno(3.20)$$
We will see that these different types of complex moduli are the ones which
appear in free fermionic models when different types of additional boundary
condition vectors are added to the ones in Eq. (3.5).

\vfill
\eject

\centerline{\bf 4. The geometric picture in $Z_2 \times Z_2$ orbifolds}

A free fermionic model with the basis ${\cal B}=\{{\bf 1},S\}$ is equivalent
to the Narain's toroidal compactification of the heterotic string[\NAR].
The 132 untwisted moduli $h_{ij}$ in Eq. (3.7) correspond to the background
fields of the toroidal compactification scheme[\MOD], i.e. the metric
$G_{\mu\nu}$,
the antisymmetric tensor $B_{\mu\nu}$ and the Wilson lines $A^a_{\mu}$,
where $\mu,\nu=1,\ldots,6$, $a=1,\ldots,16$.

Realistic superstring models in the free fermionic formulation with the basis
${\cal B^{\prime}}=\{{\bf 1},S,b_1,b_2,b_3\}$ can be seen as $Z_2 \times Z_2$
orbifolds (at the maximally symmetric point in background field
space)[\ZT]. A $Z_2 \times Z_2$ orbifold has three complex planes
corresponding to three tori $T^2$ which give the compactified six dimensions.
This orbifold has three twisted
sectors in which two complex planes are rotated (twisted) by $\pi$ and the
third one left unchanged. Each of the basis vectors $\{b_1,b_2,b_3\}$ of the
free fermionic models
gives one twist of the $Z_2 \times Z_2$ orbifold. To see this
consider the bosonic dimensions defined by bosonization
$$e^{i X_L^i}={1 \over \sqrt2}(y^i+i \omega^i), \qquad e^{i X_R^i}=
{1 \over \sqrt2}({\bar y}^i+i {\bar \omega}^i), \qquad (i=1,\ldots,6)
\eqno(4.1)$$
Then, we can form the three complex planes $Z_k^{\pm}$, $k=1,2,3$
$$Z_k^{\pm}={1\over \sqrt2}(X^{2k-1} \pm iX^{2k}), \qquad \psi_k^{\pm}=
{1\over \sqrt2}(\chi^{2k-1} \pm i\chi^{2k}),  \eqno(4.2)$$
where $X=X_L+X_R$. $\psi_k^{\pm}$ are the supersymmetric partners of the
$X_L^i$. Now, since under $b_1$
$$\eqalignno{&y^{1,2} \to -y^{1,2}, \qquad  \omega^{1,2} \to
-\omega^{1,2}, &(4.3a)\cr
             &y^{3,4,5,6} \to y^{3,4,5,6}, \qquad \omega^{3,4,5,6} \to
-\omega^{3,4,5,6}, &(4.3b)}$$
we obtain
$$e^{iX_L^{1,2}} \to -e^{iX_L^{1,2}}=e^{i(X_L^{1,2}+\pi)},
\quad e^{iX_L^{3,4,5,6}} \to e^{-iX_L^{3,4,5,6}}. \eqno(4.4)$$
Due to the world--sheet left--right symmetry of $b_1$ (and also $b_2$ and
$b_3$), the transformation of all right--handed fermions are identical to those
of the left--handed ones. As a result, $X_R^i$ transform exactly as $X_L^i$
which means that the orbifold is symmetric[\ORB]. This in turn gives
$$Z_1^{\pm} \to Z_1^{\pm} +2\pi+2\pi i, \qquad Z_{2,3}^{\pm} \to
-Z_{2,3}^{\pm}=
e^{i\pi} Z_{2,3}^{\pm}. \eqno(4.5)$$
Thus, we see that the effect of $b_1$ is to shift the first complex plane and
rotate the second and third complex planes by $\pi$. Similarly, $b_2$ shifts
the second plane and rotates the first and the third ones and $b_3$ shifts the
third plane and rotates the first and the second ones. This establishes the
connection between free fermionic strings with the basis ${\cal B^{\prime}}=
\{{\bf 1},S,b_1,b_2,b_3\}$ and $Z_2 \times Z_2$ orbifolds.
Note that $\{b_1,b_2,b_3\}$ divide the internal world--sheet fermions
($y^{1,\ldots,6}, \omega^{1,\ldots,6}$ and ${\bar y}^{1,\ldots,6},
{\bar \omega}^{1,\ldots,6})$ into three different sets $\{y^{3,4,5,6},
{\bar y}^{3,4,5,6}\}$, $\{y^{1,2},\omega^{5,6},
{\bar y}^{1,2},{\bar \omega}^{5,6}\}$,
$\{\omega^{1,2,3,4},{\bar \omega}^{1,2,3,4}\}$. Each
fermionic component of a bosonic dimension $X^i$ ($y^i,{\bar y}^i$ or
$\omega^i,{\bar \omega}^i$) appears in a different set.

It is well known that a $Z_2 \times Z_2$ orbifold has three $T$ type and three
$U$ type moduli[\IL]. This is exactly what we found for free fermionic
models (with
the basis given by Eq. (3.5)) in the previous section. Since it will be useful
for later discussions, we now show how this can be obtained in the orbifold
picture. The untwisted moduli of orbifolds are given by couplings among
$$S^{\prime}=\int d^2z (U_{ij} \partial Z_i^+ {\bar \partial} Z_j^+
+T_{ij}\partial Z_i^+ {\bar \partial} Z_j^-)+h.c., \eqno(4.6)$$
which are invariant under the orbifold twists. Here $U_{ij}$ and $T_{ij}$ are
$3 \times 3$ complex matrices which give the (36 real) moduli of the torus
compactification of type $U$ and $T$. It is easy to see that under the
$Z_k^{\pm}$ transformations
due to $\{b_1,b_2,b_3\}$ given above only the six diagonal moduli $T_{ii}=T_i$
and $U_{ii}=U_i$ ($i=1,2,3$ and there is no summation) remain invariant.
Therefore, a $Z_2 \times Z_2$ orbifold has three $T$ type and three $U$ type
complex moduli exactly as we found for the free fermionic models with the
basis ${\cal B^{\prime}}$.

\vfill
\eject

\centerline{\bf 5. Realistic free fermionic string models}

Realistic string models in the free fermionic formulation have additional
basis vectors in addition to ${\cal B}^{\prime}=\{{\bf 1},S,b_1,b_2,b_3\}$.
These additional basis vectors which are equivalent to Wilson lines in the
orbifold formulation are necessary to lower the number of chiral generations,
break the gauge group etc.. The internal world--sheet fermions are paired into
complex fermions or sigma models depending on their boundary conditions in
the additional basis vectors. For concreteness, in the following we will
consider standard--like superstring models[\SLM] as examples of realistic
string
models. All our results also apply to other realistic
models in the free fermionic formulation (such as the flipped $SU(5) \times
U(1)$ models[\REVAMP]).

Standard--like superstring models have three additional basis vectors in
addition to ${\cal B}^{\prime}=\{{\bf 1},S,b_1,b_2,b_3\}$. Three examples
of these additional basis vectors (denoted by $\alpha, \beta,\gamma$) are
given in Tables 1,2,3[\SLM]. In all cases, these vectors break the gauge
group to $SU(3)_C \times SU(2)_L \times U(1)^n \times H$ (where $H$ is
the hidden gauge group) and reduce the number of
generations to three. The vectors $\alpha, \beta,\gamma$ correspond to
an additional $Z_2^2 \times Z_4$ twist in the orbifold picture.
The $Z_4$ twist is required to reduce factors
of $SO(2n)$ in the gauge group to $SU(n) \times U(1)$.
Different models are obtained by choosing different boundary condition vectors.
In particular, for our purposes, different models can be obtained by
complexifying
different internal world--sheet fermions in the additional basis vectors
$\alpha, \beta,\gamma$ as can be seen from Tables 1,2,3. As for $b_i$, we
expect the additional basis vectors to project out some untwisted moduli (or
Abelian Thirring interactions) depending on the relevant boundary conditions.

Consider the first model defined by $\alpha, \beta,\gamma$ given by Table 1.
We notice that in this model there are no complex internal
fermions. Due to the boundary conditions chosen in $\alpha, \beta,\gamma$
all internal left and right--handed world--sheet fermions are paired into
12 sigma models. An immedeate consequence of this is left--right symmetry of
all boundary condition vectors for the internal fermions. This corresponds
to a symmetric orbifold[\ORB]. For example $\alpha$ gives
$$\eqalignno{&J_L^{1,2} \to J_L^{1,2}, \qquad J_L^{3,4,5,6} \to -J_L^{3,4,5,6},
&(5.1a)\cr
 &{\bar J}_R^{1,2} \to {\bar J}_R^{1,2}, \qquad {\bar J}_R^{3,4,5,6} \to -
{\bar J}_R^{3,4,5,6}. &(5.1b)}$$
One can easily check that $\beta$ and $\gamma$ also give left--right
symmetric boundary conditions. As
long as the boundary conditions for the left and right--handed world--sheet
currents are the same there is no way to eliminate any of the Abelian Thirring
operators since $J_L^i {\bar J}_R^i$ is always symmetric (for $Z_2$ twists).
All the
Abelian Thirring operators and therefore all the moduli which survived
the basis ${\cal B}^{\prime}=
\{{\bf 1},S,b_1,b_2,b_3\}$ given by Eq. (3.5) also survive
$\alpha,\beta,\gamma$.
As a result, a free fermionic model with no complex fermions (or 12 sigma
models) has three $T$ type and three $U$ type moduli given by Eq. (3.18)
exactly
as a $Z_2 \times Z_2$ orbifold. The Kahler potential for the moduli and the
coset space they span are given by Eqs. (3.19) and (3.20).

Now consider the model defined by the additional vectors in Table 2. We see
that in this case there is one left--handed ($\omega^2 \omega^3$) and one
right--handed (${\bar \omega}^2 {\bar \omega}^3$) complex fermion and 10
sigma operators. The complexification of world--sheet fermions
is equivalent to left--right asymmetric boundary conditions for the real
fermions which are complexified. This corresponds
to an asymmetric orbifold[\ASY]. For example,
from $\alpha$ we get the boundary conditions
$$\eqalignno{&J_L^{1,2} \to J_L^{1,2}, \qquad J_L^{3,4,5,6} \to -J_L^{3,4,5,6},
&(5.2a)\cr
 &{\bar J}_R^{1,3} \to {\bar J}_R^{1,3}, \qquad {\bar J}_R^{2,4,5,6} \to -
{\bar J}_R^{2,4,5,6}, &(5.2b)}$$
Notice that $J_L^{2,3}$ and ${\bar J}_R^{2,3}$, which are exactly the currents
which contain the complexified world--sheet fermions, transform differently.
The Abelian Thirring opeartors which are invariant under the
transformation in Eq. (5.2a,b) are
$$J^{1,2} {\bar J}^1, \quad J^{3,4} {\bar J}^4, \quad J^{5,6} {\bar J}^{5,6},
\eqno(5.3)$$
Repeating the same for $\beta$ and $\gamma$ we find that the Thirring
operators in Eq. (5.3) survive them too. The corresponding untwisted
moduli which survive are $H_1^{(1)},H_2^{(2)}$ given by Eq. (3.16a,b) and
$T_3$ and $U_3$ given by Eq. (3.18a,b).
Instead of the six complex moduli, we are left with the four given above.
The moduli which are projected out, $H_2^{(1)},H_1^{(2)}$, are
exactly the ones which contain the complexified (right--handed) world--sheet
fermions. The complex moduli $H_1^{(1)},H_2^{(2)}$ do not have a geometrical
meaning as the $T$ and $U$ moduli. The Kahler potential in this case is
given by the sum of two terms for $H_1^{(1)},H_2^{(2)}$ given by Eq. (3.17) and
two terms for $T_3,U_3$ given by Eq. (3.19). The coset space that the moduli
span is given by[\NLY]
$${\cal M}=\left({SO(2,1) \over {SO(2)}}\right)^2 \otimes {SO(2,2) \over {SO(2)
\times SO(2)}}. \eqno(5.4)$$

The same result can also be obtained by finding the massless states of
this model. From the
generalized GSO projections due to the basis vectors $\alpha, \beta,\gamma$
we find that the massless, gauge singlet states in the Neveu--Shwarz sector
are $\xi_{1,\ldots,5}$ which are given by $\chi^{12} {\bar y}^1 {\bar
\omega}^1,
\chi^{34} {\bar y}^4 {\bar \omega}^4, \chi^{56} {\bar y}^2 {\bar y}^3,
\chi^{56} {\bar y}^5 {\bar \omega}^5, \chi^{56} {\bar y}^6 {\bar \omega}^6$
acting on the Neveu--Shwarz vacuum $|0\r$[\SLM].
$\xi_{1,2,4,5}$ are the states $H_1^{(1)},H_2^{(2)},H_1^{(3)},
H_2^{(3)}$ from which the moduli $H_1^{(1)},H_2^{(2)}$, $T_3,U_3$ are built.
$\xi_3$ is not a modulus field but a gauge singlet matter field of the form
given by Eq. (3.3b).
This can also be seen from the cubic superpotential of the model which
contains the terms with $\xi_3$ but not the moduli $\xi_{1,2,4,5}$[\SLM].
Notice that the fermions making $\xi_3$ up are the pairs of the
complexified fermions of the model which make up the bosonic dimensions.
This result is quite general and can be explained as follows.

The basis ${\cal B^{\prime}}$ divides the internal fermions $(y^{1,\ldots,6},
\omega^{1,\ldots,6},{\bar y}^{1,\ldots,6},{\bar \omega}^{1,\ldots,6})$ into
three sets as we saw above. Two fermions corresponding to the same
bosonic coordinate do not
appear in the same set. On the other hand, world--sheet supersymmetry
coupled with left--right symmetry restricts the possible boundary conditions
severely. This can be seen from the form of $T_F$ given by Eq. (2.1). One finds
that for $T_F$ to have a well-defined boundary condition,
$$(b(\chi^i),b(y^i),b(\omega^i))= (1,1,1) \quad \hbox{or} \quad (1,0,0)
\quad \hbox{or its permutations} \eqno(5.5)$$
when $b(\psi^{\mu})=1$ and
$$(b(\chi^i),b(y^i),b(\omega^i))=(0,0,0) \quad \hbox{or} \quad (1,1,0)
\quad \hbox{or its permutations} \eqno(5.6)$$
when $b(\psi^{\mu})=0$. Together with left--right
symmetry this constrains the possible boundary conditions for the right--handed
fermions. Finally, the gauge singlets from the Neveu--Shwarz sector have
a unique form as seen from Eq. (3.14). They are made up of one complex
$\chi^{i,i+1}$ and two right--handed fermions from two different sets
(defined by ${\cal B^{\prime}}$ which are also different from the set
corresponding to $\chi^{i,i+1}$).
Using this information, it is easy to show that the gauge singlets
which survive the GSO projections due to the additional basis vectors are
of the form $\chi^{i,i+1}
{\bar y}^j {\bar \omega^j}$ ($j=i,i+1$) if ${\bar y}^j, {\bar \omega^j}$ are
not
complexified and $\chi^{i,i+1} {\bar y}^k {\bar \omega^l}$ if ${\bar y}^l,
{\bar \omega^k}$ ($i,k,l$ belong to different sets) are complexified. For
each pair (left+right handed) of complex fermions there will be
one such gauge singlet in the Neveu--Shwarz sector which is not a modulus.

The results obtained above are completely general. When there are no complex
internal fermions, all boundary conditions for the internal fermions are
left--right symmetric. As a result, all Abelian Thirring interactions are
invariant and no moduli are projected out.
When a pair of right--handed (and therefore a pair
of left--handed) internal world--sheet fermions are complexified,
the moduli or Thirring interactions containing them are projected out and
therefore the number of moduli is reduced by two. The number of moduli
is given by the number of sigma models minus six or by six minus the number of
complex left and right--handed internal fermions (since the number of complex
fermions plus the number of sigma models arising from 24 real internal
world--sheet fermions is 12). The moduli which are projected out (in a given
sector or complex plane, $j$) are of the type $H_i^{(j)}$ instead of
either $T_j$ or $U_j$ type. The result of a complexified real, internal
world--sheet fermion of of a complex plane $j$ is the elimination of the $T_j$
and $U_j$ moduli in a manner that one complex modulus of
the type $H_i^{(j)}$ remains. We will see in Section 7 that in general there
are other possibilities for the remaining moduli which depend on the details
of the left--right asymmetry of the boundary conditions for the internal
fermions. Note that due to the particular division of
the internal fermions by the basis ${\cal B^{\prime}}$ each complex fermion
contains two real fermions from two different complex planes.
As a result, when there is
only one pair (of each chirality) of complex fermions, $T_j,U_j$ are moded
out in two planes and one modulus of type $H_i^{(j)}$ remains in each plane.
When there are two complex fermions, there is one plane without moduli at all
and two planes with only one modulus of type $H_i^{(j)}$. Finally, when there
are three complex fermions, there are no moduli left. In addition, for each
pair of complex (left and right--handed) fermion there is a massless gauge
singlet which is not a modulus and which is given by
the rule in the preceding paragraph.

Consider now the model defined by the basis vectors in Table 3. We see that
in this model there are three complex left-handed ($y^3 y^6, y^1 \omega^5,
\omega^2 \omega^4$) and three complex right--handed (${\bar y}^3 {\bar y}^6,
{\bar y}^1 {\bar \omega}^5, {\bar \omega}^2 {\bar \omega}^4$) fermions.
Since the total number of (left+right--handed) complex fermions is six,
from the above discussion we expect that all untwisted moduli given by
Eq. (3.13) are projected out. For example from $\alpha$ we find
$$J^{1,\ldots,6} \to -J^{1,\ldots,6}, \qquad {\bar J}^{1,\ldots,6} \to
{\bar J}^{1,\ldots,6}, \eqno(5.7)$$
and as a result all untwisted moduli (or Abelian Thirring interactions) are
projected out. The same result is also obtained from $\beta$ and $\gamma$
separately.
One can verify this by finding the massless spectrum of the
model and checking that all untwisted moduli fields are eliminated by the GSO
projections due to $\alpha,\beta,\gamma$. The massless gauge neutral states
(which are not moduli) are given by $\chi^{12} {\bar \omega}^3 {\bar \omega}^6,
\chi^{34} {\bar y}^5 {\bar \omega}^1,\chi^{56} {\bar y}^2 {\bar y}^4$ acting
on the Neveu--Shwarz vacuum $|0\r$[\SLM]. This is exactly what one obtains
according to the rule we gave above since, in this model, the complexified
fermions are
${\bar y}^3 {\bar y}^6, {\bar y}^1 {\bar \omega}^5, {\bar \omega}^2
{\bar \omega}^4$. These fields appear in the cubic superpotential of the
model which confirms that they are not moduli[\SLM].

\vfill
\eject

\centerline{\bf 6. Geometric picture of the Wilson lines}

In this section we find the geometrical meaning of the additional basis
vectors $\alpha,\beta,\gamma$ of realistic free fermionic models.  We
translate the boundary conditions of the internal fermions in the free
fermionic models into transformations of the three
complex planes in the orbifold picture by using the connection to the orbifold
picture described in Section 4. Among untwisted moduli given by the action
[\ORBM]
$$S^{\prime}=\int d^2z (U_i \partial Z_i^+ {\bar \partial} Z_i^+ +T_i\partial
Z_i^+ {\bar \partial} Z_i^-)+h.c., \eqno(6.1)$$
only those which remain invariant under the transformations of the complex
planes due to the additional basis vectors survive. First, we consider the
model described by
the basis vectors of Table 1. It is easy to show that from $\alpha$
$$\eqalignno{&y^1+i \omega^1 \to -y^1-i \omega^1, \qquad y^2+i \omega^2 \to
-y^2-i \omega^2, &(6.2a)\cr
&{\bar y}^1+i {\bar \omega}^1 \to -{\bar y}^1-i {\bar \omega}^1,
\qquad {\bar y}^2+i {\bar \omega}^2 \to
-{\bar y}^2-i {\bar \omega}^2, &(6.2b)}$$
and using the bosonization formulas of Eq. (4.1)
$$X_L^{1,2} \to X_L^{1,2}+\pi, \qquad X_R^{1,2} \to X_R^{1,2}+\pi, \eqno(6.3)$$
and therefore
$$Z_1^{\pm} \to Z_1^{\pm}+2\pi+2\pi i. \eqno(6.4)$$
Obviously this does not eliminate any moduli appearing in Eq. (3.13) since
the Thirring interactions contain only derivatives of $Z_k^{\pm}$. Repeating
this for the second complex plane we get
$$\eqalignno{&y^3+i \omega^3 \to y^3-i \omega^3, \qquad y^4+i \omega^4 \to
-y^4+i \omega^2, &(6.5a)\cr
&{\bar y}^3+i {\bar \omega}^3 \to {\bar y}^3-i {\bar \omega}^3,
\qquad {\bar y}^4+i {\bar \omega}^4 \to
-{\bar y}^4+i {\bar \omega}^4, &(6.5b)}$$
which gives
$$X^3_{L,R} \to -X^3_{L,R}, \qquad X^4_{L,R} \to -X^4+2\pi, \eqno(6.6)$$
and therefore
$$Z_2^{\pm} \to e^{i\pi} Z_2^{\pm}+2\pi i. \eqno(6.7)$$
This does not eliminate any moduli either since in Eq. (6.1) $Z_k^{\pm}$
always come in pairs. The same is true for the third complex plane for which
$\alpha$ gives
$$Z_3^{\pm} \to e^{i\pi} Z_3^{\pm} +2\pi. \eqno(6.8)$$
One can easily check that the other basis vectors $\beta$ and $\gamma$ also
result in tranformations of the three complex planes which are combinations
of multiplication by $e^{i\pi}$ and/or addition of $2\pi$ and/or $2\pi i$,
i.e. rotations by $\pi$ and complex translations.
This follows from the fact that free fermionic models correspond to $Z_2
\times Z_2$ orbifolds and in this model all boundary conditions are
world--sheet left--right symmetric. None of these tranformations of complex
planes can eliminate any moduli due to the form of the moduli action in
Eq. (6.1).

Now, consider the model defined by the basis vectors in Table 2. From the
previous section, we know that in this case there are only four moduli. This,
as we saw, is due to the complexified fermions which are equivalent to
left--right asymmetric boundary conditions. We find from $\alpha$
$$\eqalignno{&y^1+i \omega^1 \to -y^1-i \omega^1, \qquad y^2+i \omega^2 \to
-y^2-i \omega^2, &(6.9a)\cr
&{\bar y}^1+i {\bar \omega}^1 \to -{\bar y}^1-i {\bar \omega}^1,
\qquad {\bar y}^2+i {\bar \omega}^2 \to
-{\bar y}^2+i {\bar \omega}^2, &(6.9b)}$$
which gives by bosonization
$$X^1 \to X^1+2\pi, \qquad X^2 \to X^2_L-X^2_R+2\pi. \eqno(6.10)$$
For the second complex plane we have similarly
$$\eqalignno{&y^3+i \omega^3 \to y^3-i \omega^1, \qquad y^4+i \omega^4 \to
-y^2+i \omega^2, &(6.11a)\cr
&{\bar y}^3+i {\bar \omega}^3 \to {\bar y}^3+i {\bar \omega}^3,
\qquad {\bar y}^4+i {\bar \omega}^4 \to
-{\bar y}^4+i {\bar \omega}^4, &(6.11b)}$$
which gives from Eq. (4.1)
$$X^3 \to X^3_R-X^3_L, \qquad X^4 \to -X^4+2\pi. \eqno(6.12)$$
These are not geometrical transformations of the compactified bosonic
dimensions since $X^2_L$, $X^2_R$ and $X^3_L$, $X^3_R$ transform differently.
This is a result of
the fact that the complexified fermions are equivalent to left--right
asymmetric
boundary conditions in $\alpha$. In fact, the left--right asymmetric
transformations occur in the dimensions whose fermionic parts were
complexified, i.e. $X^{2,3}$ where $\omega^{2,3},{\bar \omega}^{2,3}$ are
complexified. It can be easily shown that in this case the third complex
plane has a geometrical transformation
$$Z_3^{\pm} \to e^{i\pi} Z_3^{\pm} +2\pi, \eqno(6.13)$$
as a result of the left--right symmetric boundary conditions in this plane.
Thus, we find that complexified world--sheet fermions result in left--right
asymmetric transformations of their corresponding bosonized dimensions.
The untwisted moduli action Eq. (6.1) is not invariant under the
transformations
given by Eqs. (6.10) and (6.12) which means that at least some moduli are
projected out by $\alpha$ in this model.
By writing out Eq. (6.1) in terms of $X^i_{L,R}$ one can show that in this case
the moduli action is invariant under Eqs. (6.10) and (6.12) only if
$T_1=U_1$ and $T_2=-U_2$ respectively. From the definition of the $T$ and $U$
moduli in Eq. (3.18) in terms
of $H_j^{(i)}$ we find that this corresponds to $H_2^{(1)}=H_1^{(2)}=0$.
Another way of seeing this is to notice that under the transformations in
Eq. (6.10) and (6.12) the moduli behave as $T_1 \to U_1$, $U_1 \to T_1$,
and $T_2 \to -U_2$, $U_2 \to -T_2$ (or $H_2^{(1)} \to -H_2^{(1)}$,
$H_1^{(2)} \to -H_1^{(2)}$, $H_1^{(1)} \to H_1^{(1)}$ and $H_2^{(2)} \to
H_2^{(2)}$) which gives the same result.
Thus the transformations due to $\alpha$ result in the elimination of
exactly the two moduli that we found
in the previous section. Repeating the same exercise for $\beta$ and $\gamma$
we find that only the moduli $H_2^{(1)},H_1^{(2)}$ are eliminated in this
model. From Eq. (3.18) it is easy to see that when $T_i=U_i$ or $T_i=-U_i$
the remaining modulus is $H^{(i)}_1$ or $H^{(i)}_2$ respectively. As
mentioned before, we will see in Section 7 that there are other possibilities
for the remaining moduli. These result from different left--right
asymmetric boundary conditions for the internal fermions.

Using the bosonization formulas, we translated the boundary conditions of the
internal fermions into transformations on the complex planes of the
$Z_2 \times Z_2$ orbifold. We found that as long as there are no complex
internal fermions, i.e. all boundary conditions are left--right symmetric,
the complex planes are either shifted by a complex constant or multiplied
by $e^{i \pi}$. The moduli action is symmetric under these and therefore
no moduli are projected out in this case. When there is a complex fermion,
i.e. left--right asymmetric boundary conditions, two bosonized dimensions
in two different complex planes transform in a left--right asymmetric manner.
In this case, the moduli action is invariant only if the moduli of these
planes obey certain conditions such as $T=U$ or $T=-U$. These conditions
are exactly the ones which eliminate the moduli which are projected out
in the free fermionic picture. It is easy to see that we obtain exactly the
same number and type of moduli when there are one, two or three complex
fermions as in the free fermionic formulation.

Now consider the model with the basis vectors of Table 3. In the previous
section, we found that in this model all moduli are projected out as a result
of the six complex left and right--handed fermions. Translating the boundary
conditions in $\alpha$ into transformations of $Z_k^{\pm}$ as above we find
$$\eqalignno{&X^1 \to X^1_L-X^1_R+\pi, \qquad X^2 \to X^2_L-X^2_R+\pi,
&(6.14a,b)\cr
&X^3 \to X^3_L-X^3_R+\pi, \qquad X^4 \to X^4_L-X^4_R+2\pi, &(6.14c,d)\cr
&X^5 \to X^5_L-X^5_R+2\pi, \qquad X^6 \to X^6_L-X^6_R+\pi. &(6.14e,f)}$$
In this case, due to the three complex fermions in $\alpha$ there are
left--right
asymmetric transformations in all the complex planes. Note that for all
bosonic dimensions
$X^i \to X^i_L-X^i_R+const$ which is different from the previous case in which
there was only one bosonic coordinate transforming asymmetrically in the two
complex planes. Due to the transformations in Eq. (6.14) all the moduli $T_i$
and $U_i$ are projected out since none of the terms in the moduli action,
Eq. (6.1) are invariant. This can be easily seen as follows. The moduli action
is invariant under the transformations in Eq. (6.14a,c,e) only if
$T_i=U_i$ but invariant under those in Eq. (6.14b,d,f) only
if $T_i=-U_i$. When both act simulataneously, there are no invariant Thirring
interactions in the moduli action. Another way of seeing this is to notice
from Eq. (3.18) that under the transformations in Eq. (6.14), $T_i \to U_i$ and
$U_i \to -T_i$ which means that for the moduli action to be invariant all
moduli must vanish.
This is exactly what we found in the previous
section from the boundary conditions of Abelian Thirring interactions.
The same result can also be obtained from $\beta$ and $\gamma$ separately.

\bigskip
\centerline{\bf 7. General Results}

In previous sections, we investigated three different
standard--like superstring models and saw how the boundary conditions for
the internal fermions determine the number and type of
moduli in a given model.
In this section, we investigate all possible transformations of the complex
planes $Z_k^{\pm}$ which arise from all possible boundary conditions for
the internal fermions. We consider
free fermionic models with only $Z_2$ twists for the internal fermions
since all realistic string models constructed so far are of this type.
Our aim is to show that the rule we gave in Section
5 for finding the number of untwisted moduli is valid in general and to find
all the possibilities for the type of remaining moduli.

We first list all boundary conditions which are left--right symmetric.
There are four possible boundary conditions for $y^i+i \omega^i$ which
is equivalent to the bosonic dimension $X^i$:
$$\eqalignno{&y^i+i \omega^i \to -y^i+i \omega^i, &(7.1a)\cr
             &y^i+i \omega^i \to -y^i- i \omega^i, &(7.1b)\cr
             &y^i+i \omega^i \to y^i- i \omega^i, &(7.1c)\cr
             &y^i+i \omega^i \to y^i+ i \omega^i. &(7.1d)}$$
Due to the left--right symmetry, the right--handed fermions, ${\bar y}^i,
{\bar \omega}^i$,
transform in the same way as the left--handed ones.
Each of the two bosonic coordinates in the complex planes $Z_k^{\pm}$
transforms in one of the four ways given above. Thus, overall there are  16
different transformations for $Z_k^{\pm}$ when the boundary conditions for
world--sheet fermions are left--right symmetric. These transformations
are listed in Table 4.
We see that eight of the transformations take the complex plane $Z_k^{\pm}$
to itself (modulo multiplication by $e^{i\pi}$ and addition of constants)
whereas eight others take it to its complex conjugate, i.e.
$Z_k^{\pm} \to Z_k^{\mp}$. We have seen that, because of the form of the moduli
action, multiplication by $e^{i\pi}$ and addition of constants to $Z_k^{\pm}$
do not change the action. Therefore no moduli are projected out as a result of
the former eight transformations. On the other hand, for the eight
transformations which result in $Z_k^{\pm} \to Z_k^{\mp}$
(modulo multiplication by $e^{i\pi}$ and addition of constants), the moduli
action is invariant only if $T_i = {\bar T_i}$ or $U_i = {\bar U_i}$. From the
definitions of the $T$ and $U$ moduli we find that in this case the complex
moduli of the type $H_i^{(2)}$ are projected out. This is contrary to the
rules we gave in Section 5
since there we concluded that moduli are projected out only
as a result of complex world--sheet fermions which are equivalent to
left--right asymmetric boundary conditions.

We now demonstrate that world--sheet supersymmetry does not allow
transformations of the type $Z_k^{\pm} \to Z_k^{\mp}$.
In order to get a transformation of the type $Z_k^{\pm} \to Z_k^{\mp}$,
the two real bosonic coordinates $X^{2k-1}$ and $X^{2k}$ of a given complex
plane $Z_k^{\pm}$ must transform in
opposite ways, i.e. $X^{2k-1} \to \pm X^{2k-1}$ and $X^{2k} \to \mp X^{2k}$.
This means that the corrsponding fermionic degrees of freedom must
transform as
$$\eqalignno{&y^{2k-1}+i \omega^{2k-1} \to y^{2k-1}-i \omega^{2k-1}, &(7.2a)\cr
         &y^{2k}+i \omega^{2k} \to y^{2k}+i \omega^{2k}, &(7.2b)}$$
or as
$$\eqalignno{&y^{2k-1}+i \omega^{2k-1} \to y^{2k-1}+i \omega^{2k-1}, &(7.3a)\cr
         &y^{2k}+i \omega^{2k} \to y^{2k}-i \omega^{2k}, &(7.3b)}$$
up to overall signs (which are equivalent to a factor $e^{i\pi}$) and
constants which are not relevant for our discussion. The right--handed
world--sheet fermions transform the same way due to the assumed left--right
symmetry of the boundary conditions. In short, this requires for both
left and right--handed fermions,
that $\omega^{2k-1}$ transforms opposite to $y^{2k-1}$ whereas $\omega^{2k}$
transforms as $y^{2k}$ or vice versa.
On the other hand, the world--sheet supersymmetry which is generated by the
world-sheet supercurrent given by Eq. (2.1) restricts the possible boundary
conditions severely. From Eq. (2.1) we see that in order for $T_F$ to have a
well--defined boundary condition one must have in every basis vector $b$
$$(b(\chi^i),b(y^i),b(\omega^i))= (1,1,1) \quad \hbox{or} \quad (1,0,0)
\quad \hbox{or its permutations} \eqno(7.4)$$
when $b(\psi^{\mu})=1$ and
$$(b(\chi^i),b(y^i),b(\omega^i))=(0,0,0) \quad \hbox {or} \quad (1,1,0)
\quad \hbox {or its permutations} \eqno(7.5)$$
when $b(\psi^{\mu})=0$. Using the fact that for any complex
plane, the two corresponding $\chi^{2k-1}$ and $\chi^{2k}$ have the same
boundary conditions (since they are always complexified) we find that
both $y^{2k-1}$ and $y^{2k}$ transform either as or opposite to $\omega^{2k-1}$
and $\omega^{2k}$ respectively. In other words, if
$X^{2k-1} \to \pm X^{2k-1}$ then $X^{2k} \to \pm X^{2k}$ necessarily due
to world--sheet supersymmetry.
Therefore, we conclude that world--sheet supersymmetry does not allow the
boundary conditions which correspond to complex conjugation of the complex
planes. As a result, basis vectors with left--right symmetry correspond
only to tranformations with multiplication by $e^{i\pi}$ and addition of
constants, i.e. rotations by $\pi$ and translations of the complex planes.
Under these transformations the moduli action is invariant and
no moduli are projected out as we found before.

We now consider the left--right asymmetric boundary conditions. We saw
above that in this case there are complex world--sheet fermions and the
corresponding left--right asymmetry results in the elimination of moduli.
There are four possible
transformations of the bosonic coordinates when left--right asymmetry is
included:
$$\eqalignno{&X^i \to X^i, \qquad X^i \to -X^i, &(7.6a,b)\cr
         &X^i \to X^i_L-X^i_R, \qquad X^i \to X^i_R-X^i_L. &(7.6c,d)}$$
Combinations of the first two in Eq. (7.6a,b) are already included in the
previous discussion
since they correspond to the left--right symmetric case. Then, we are left with
12 possibilities. The moduli action in Eq. (6.1) is not invariant under any of
these remaining transformations. The eight tranformations given in
Table 5 can be a symmetry only if the conditions specified in the table
are satisfied. These are the cases in which only one bosonic coordinate
has a left--right asymmetric transformation. Note that if the first
dimension of the complex plane $i$ is symmetric (asymmetric)
and the second one goes to $L-R$ ($R-L$) the condition is $T_i=U_i$. If the two
bosonic coordinates are interchanged with respect
to the previous case then the condition becomes $T_i=-U_i$.
As we saw before, instead of the two complex moduli $T_i$ and $U_i$
corresponding to the complex plane we are left with one complex modulus
$H_1^{(i)}$ ($H_2^{(i)}$) in the former (latter) case. This is
easily seen by inspecting the definitions
of the moduli in Eq. (3.18) and the necessary conditions from Table 5.
The three string models we considered in Section 5 are examples of this kind.
In the remaining cases, when the first dimension of the complex plane $i$ is
symmetric (asymmetric) and the second one goes to $R-L$ ($L-R$)
the condition is $T_i={\bar U}_i$. If the two bosonic coordinates are
interchanged then the condition becomes $T_i=- {\bar U}_i$. Now, from
Eqs. (3.18) and from the definition of the moduli, we find that instead
of one complex moduli of the type $H_{1,2}^{(i)}$
we are left with two real ones of the type $h_{ij}$. In the former case,
these are $h_{i1}$ and $h_{i2}$ whereas in the latter case they are
$h_{2i}$ and $h_{2i+1}$.

These are all the possibilities for the surviving untwisted moduli when there
are complex internal fermions (in models with
only $Z_2$ twists for the internal fermions).
We see that the type of moduli present depend on the
details of the left--right asymmetry in the boundary conditions for the
internal fermions. We find that every
complexified world--sheet fermion results in a left--right asymmetric
transformation of a bosonic coordinate which in turn results in a condition
on the moduli. Since each condition eliminates one complex
modulus, we obtain the simple rule given in Section 5, namely that each
pair of complex (left and right--handed) fermions eliminates two complex
(or four real) moduli.

When there are two fermions in a given complex plane with left--right
asymmetric boundary conditions, both bosonic coordinates of the complex
plane transform in a left--right asymmetric manner. Then, two conditions of
the type given in Table 5 must be satisfied simultaneously. From Table
5 we see that there are two possibilities: either both bosonic dimensions
transform in the same
left--right asymmetric way or in opposite ways. From inspecting the required
conditions on the moduli for each case in Table 5, we see that in the former
case the two conditions must be different.
Since two different conditions cannot be satisfied simultaneously, both
moduli of $T$ and $U$ type corresponding to the complex plane are
projected out. Thus, if there are two complexified fermions in a given plane
all moduli corresponding to the plane are eliminated. (Note that complexified
world--sheet fermion pairs always contain fermions from different complex
planes. This is due to the particular division of the internal fermions
by the basis ${\cal B^{\prime}}$ as mentioned in Section 2. Then, two fermions
from the
same plane are complexified only when there are four or six complexified
fermions of each chirality.) This is exactly the
result obtained from the rule given in Section 5. As we saw,
this happens in our third example in which case there are
two such fermions in each complex plane and therefore all moduli are moded out.

In the latter case, in which the bosonic dimensions transform in opposite
ways, a priori the two conditions on the moduli are not mutually exclusive.
At first sight, one may think that in this case there will be some moduli
left even with two complexified fermions in a given plane. This violates the
rules we gave in Section 5. Once again we can
show that world--sheet supersymmetry does not allow these transformations.
Consider the case where
$$X^1 \to X^1_L-X^1_R, \qquad X^1 \to X^2_R-X^2_L, \eqno(7.7a,b)$$
which corresponds to the boundary conditions for the fermions given by
$$\eqalignno{&y^1+i \omega^1 \to y^1+i \omega^1, \qquad y^2+i \omega^2 \to
y^2-i \omega^2, &(7.8a)\cr
&{\bar y}^1+i {\bar \omega}^1 \to {\bar y}^1-i {\bar \omega}^1,
\qquad {\bar y}^2+i {\bar \omega}^2 \to
{\bar y}^2+i {\bar \omega}^2. &(7.8b)}$$
{}From the constraints on the boundary conditions coming from the supercurrent,
i.e. Eqs. (7.4) and (7.5),
we find that the above is not possible. The same result holds for the case
in which $X^1$ and $X^2$ are interchanged.
This proves our results for the number and type of untwisted moduli for free
fermionic models with only $Z_2$ twists for the internal fermions.

\bigskip
\centerline{\bf 7. Conclusions}

In this paper, we have investigated the untwisted moduli of realistic free
fermionic
string models. The untwisted moduli correspond to the exactly marginal
operators of the conformal field theory i.e. free fermions on the
world--sheet. These are given by the Abelian Thirring interactions of the form
$J_L^i {\bar J}_R^j$. The physical untwisted moduli correspond to those
Thirring interactions which
are compatible with the boundary conditions arising from the basis vectors
of the model. This in turn means that the number and type of untwisted
moduli of free fermionic strings are determined by the basis vectors
which define the model.

We found that when the boundary conditions for the internal fermions are
left--right symmetric, i.e. there are no complex internal fermions,
no moduli are projected out. On the other hand, when there are complex internal
fermions, i.e. left--right asymmetric boundary conditions for the internal
fermions, each complex (left or right--handed) internal fermion
reduces the number of moduli by one. Since complex fermions come always in
left--right symmetric pairs, the number of moduli is reduced by two for each
pair of complex internal fermions. Equivalently,
each complexified real internal fermion of a given chirality reduces
the number of moduli by one. This is because a complex fermion is equivalent to
left--right asymmetric boundary conditions for the two real fermions which
make it up. When this happens two left or right--handed currents
($J^i_L$ or ${\bar J}^j_R$) get a twist and the Abelian Thirring operators
which contain them are projected out.
Depending on the boundary conditions in the basis vectors of the model, the
internal fermions are either complexified or paired into sigma models.
The number of moduli is given by the number
of sigma operators of the model minus six or the six minus the number of
complex (left+ right--handed) fermions. Whenever a fermion belonging to a
complex plane $i$ is complexified, the $T_i$ and $U_i$ moduli corresponding
to the plane are projected out. Then, there remains only one complex modulus
of the type $H_j^{(i)}$ or two real ones of the type $h_{ij}$ depending on
the left--right asymmetric boundary conditions as given in Table 5. In
this manner, the boundary conditions for the internal fermions in the basis
vectors determine the number and type of untwisted moduli of the string model.
Due to the particular division of the internal
fermions by the basis ${\cal B^{\prime}}$ into three different sets, each
complex fermion contains two fermions from two different complex planes.
As a result, when there is only one pair (of each chirality) of complex
internal fermions, $T_i,U_i$ are moded
out in two planes and one modulus of type $H_j^{(i)}$ or two real moduli
of type $h_{ij}$ remain in each of the two planes.
When there are two complex internal fermions, there is one plane without
moduli at all and two planes with only one modulus of type $H_j^{(i)}$ or
two real moduli of type $h_{ij}$. Finally, when there are three complex
internal fermions, there are no moduli left. In addition, for each
pair of complex internal fermions there is a massless gauge singlet which
is not a modulus in the Neveu--Shwarz sector.

We also gave a geometrical interpretation of our above results, by
using the equivalence of free fermionic models to $Z_2 \times Z_2$ orbifolds
with asymmetric Wilson lines. Using the bosonization formulas, one can
translate the boundary conditions of the internal fermions into transformations
of the complex planes of the orbifold. We found that as long as there are no
complex internal fermions, i.e. all boundary conditions for the internal
fermions are left--right symmetric,
the complex planes are either shifted by a complex constant (translation) or
multiplied by $e^{i \pi}$ (rotation by $\pi$). The moduli action is
symmetric under these transformations and therefore no moduli are projected
out in this case. When there is a complex internal fermion,
i.e. left--right asymmetric boundary conditions, two bosonized dimensions
in two different complex planes transform in a left--right asymmetric manner.
In this case the moduli action is invariant only if the moduli of these
planes obey certain conditions as given in Table 5. These conditions
are exactly the ones which eliminate the moduli which are projected out
in the free fermionic picture. Depending on the properties of the left--right
asymmetric transformation (or boundary conditions) one is left with either
a complex modulus $H_j^{(i)}$ or two real ones $h_{ij}$. The cases with two
and three complex fermions can also be understood in a similar way which gives
results identical to those in the free fermionic formulation.

We have proved our results by considering all possible boundary conditions
corresponding to $Z_2$ twists for the internal fermions. We showed that the
boundary conditions for the internal fermions or
transformations of the complex planes of the orbifold which violate our rules
are not allowed by world--sheet supersymmetry. When there is left--right
symmetry (no complex internal fermions) there are transformations which take
a complex plane to its complex conjugate (see Table 4) and which violate our
results. These are not allowed because
in this case the boundary condition of the world--sheet supercurrent $T_F$
is not well-defined. Similarly when there are two complex internal fermions in
a given plane with the opposite left--right asymmetry, our results are
violated. Similarly to the previous case, these boundary conditions (or
transformations) are not allowed due to world--sheet supersymmetry. We
conclude that our results are valid in general for the realistic free
fermionic models with $Z_2$ twists for the internal fermions.

\bigskip
\centerline{\bf Acknowledgements}
I thank Alon Faraggi for usefull discussions during the early stages of this
work.
This work is supported by the Department of Particle Physics and a Feinberg
Fellowship.

\refout
\vfill
\eject

\end
\bye

\input tables.tex
\nopagenumbers
\magnification=1000
\baselineskip=18pt
\hbox
{\hfill
{\begintable
\  \ \|\ ${\psi^\mu}$ \ \|\ $\{{\chi^{12};\chi^{34};\chi^{56}}\}$\ \|\
{${\bar\psi}^1$, ${\bar\psi}^2$, ${\bar\psi}^3$,
${\bar\psi}^4$, ${\bar\psi}^5$, ${\bar\eta}^1$,
${\bar\eta}^2$, ${\bar\eta}^3$} \ \|\
{${\bar\phi}^1$, ${\bar\phi}^2$, ${\bar\phi}^3$, ${\bar\phi}^4$,
${\bar\phi}^5$, ${\bar\phi}^6$, ${\bar\phi}^7$, ${\bar\phi}^8$} \crthick
$\alpha$
\|\ 1 \|
$\{1,~0,~0\}$ \|
1, ~~1, ~~1, ~~0, ~~0, ~~1 ,~~0, ~~0 \|
1, ~~1, ~~0, ~~0, ~~0, ~~0, ~~0, ~~0 \nr
$\beta$
\|\ 1 \| $\{0,~1,~0\}$ \|
1, ~~1, ~~1, ~~0, ~~0, ~~0, ~~1, ~~0 \|
1, ~~1, ~~0, ~~0, ~~0, ~~0, ~~0, ~~0 \nr
$\gamma$
\|\ 1 \|
$\{0,~0,~1\}$ \|
{}~~$1\over2$, ~~$1\over2$, ~~$1\over2$, ~~$1\over2$,
{}~~$1\over2$, ~~$1\over2$, ~~$1\over2$, ~~$1\over2$ \| $1\over2$, ~~$1\over2$,
{}~~$1\over2$, ~~$1\over2$, ~~1,
{}~~0, ~~0, ~~0 \endtable}
\hfill}
\smallskip
{\hfill
{\begintable
\  \ \|\
${y^3 {\bar y}^3}$,  ${y^4{\bar y}^4}$, ${y^5{\bar y}^5}$,
${y^6{\bar y}^6}$
\ \|\ ${y^1 {\bar y}^1}$,  ${y^2{\bar y}^2}$,
${\omega^5{\bar\omega}^5}$,
${\omega^6{\bar\omega}^6}$
\ \|\ ${\omega^1{\bar \omega}^1}$,  ${\omega^2{\bar \omega}^2}$,
${\omega^3{\bar \omega}^3}$,  ${\omega^4{\bar \omega}^4}$  \crthick
$\alpha$ \|
1, ~~~0, ~~~~0, ~~~~1 \|
0, ~~~0, ~~~~1, ~~~~0 \|
0, ~~~0, ~~~~0, ~~~~1 \nr
$\beta$ \|
0, ~~~0, ~~~~0, ~~~~1 \|
0, ~~~1, ~~~~1, ~~~~0 \|
1, ~~~0, ~~~~0, ~~~~0 \nr
$\gamma$ \|
1, ~~~1, ~~~~0, ~~~~0 \|\
1, ~~~0, ~~~~0, ~~~~0 \|
0, ~~~1, ~~~~0, ~~~~0  \endtable}
\hfill}
\smallskip
\parindent=0pt
\hangindent=39pt\hangafter=1
\baselineskip=18pt

{{\it Table 1.} A three generations ${SU(3)\times SU(2)\times U(1)^2}$
model. The choice of generalized GSO coefficients is:
${c\left(\matrix{b_j\cr
                                    \alpha,\beta,\gamma\cr}\right)=
-c\left(\matrix{\alpha\cr
                                    1\cr}\right)=
c\left(\matrix{\alpha\cr
                                    \beta\cr}\right)=
-c\left(\matrix{\beta\cr
                                    1\cr}\right)=
c\left(\matrix{\gamma\cr
                                    1,\alpha\cr}\right)=
-c\left(\matrix{\gamma\cr
                                    \beta\cr}\right)=
-1}$ (j=1,2,3),
with the others specified by modular invariance and space--time
supersymmetry.  The $16$ right--moving
internal fermionic states
$\{{\bar\psi}^{1,\cdots,5},{\bar\eta}^1,
{\bar\eta}^2,{\bar\eta}^3,{\bar\phi}^{1,\cdots,8}\}$,
correspond to the $16$ dimensional compactified  torus of the ten dimensional
heterotic string.
The 12 left--moving and 12 right--moving real internal fermionic states
correspond to the six left
and six right compactified dimensions in the bosonic language.}
\vskip 2.5cm

\vfill
\eject

\end
\bye

\input tables.tex
\nopagenumbers
\magnification=1000
\baselineskip=18pt
\hbox
{\hfill
{\begintable
\  \ \|\ ${\psi^\mu}$ \ \|\ $\{{\chi^{12};\chi^{34};\chi^{56}}\}$\ \|\
{${\bar\psi}^1$, ${\bar\psi}^2$, ${\bar\psi}^3$,
${\bar\psi}^4$, ${\bar\psi}^5$, ${\bar\eta}^1$,
${\bar\eta}^2$, ${\bar\eta}^3$} \ \|\
{${\bar\phi}^1$, ${\bar\phi}^2$, ${\bar\phi}^3$, ${\bar\phi}^4$,
${\bar\phi}^5$, ${\bar\phi}^6$, ${\bar\phi}^7$, ${\bar\phi}^8$} \crthick
$\alpha$
\|\ 1 \|
$\{1,~0,~0\}$ \|
1, ~~1, ~~1, ~~0, ~~0, ~~1 ,~~0, ~~1 \|
1, ~~1, ~~1, ~~1, ~~0, ~~0, ~~0, ~~0 \nr
$\beta$
\|\ 1 \| $\{0,~1,~0\}$ \|
1, ~~1, ~~1, ~~0, ~~0, ~~0, ~~1, ~~1 \|
1, ~~1, ~~1, ~~1, ~~0, ~~0, ~~0, ~~0 \nr
$\gamma$
\|\ 1 \|
$\{0,~0,~1\}$ \|
{}~~$1\over2$, ~~$1\over2$, ~~$1\over2$, ~~$1\over2$,
{}~~$1\over2$, ~~$1\over2$, ~~$1\over2$, ~~$1\over2$ \| $1\over2$, ~~0, ~~1,
{}~~1, ~~$1\over2$, ~~$1\over2$, ~~$1\over2$,
{}~~0  \endtable}
\hfill}
\smallskip
{\hfill
{\begintable
\  \ \|\
${y^3 {\bar y}^3}$,  ${y^4{\bar y}^4}$, ${y^5{\bar y}^5}$,
${y^6{\bar y}^6}$
\ \|\ ${y^1 {\bar y}^1}$,  ${y^2{\bar y}^2}$,
${\omega^5{\bar\omega}^5}$,
${\omega^6{\bar\omega}^6}$
\ \|\ ${\omega^2{\omega}^3}$,  ${\omega^1{\bar\omega}^1}$,
${\omega^4{\bar\omega}^4}$,  ${\bar \omega}^2{\bar\omega}^3$  \crthick
$\alpha$ \|
1, ~~~0, ~~~~0, ~~~~1 \|
0, ~~~0, ~~~~1, ~~~~0 \|
0, ~~~0, ~~~~1, ~~~~1 \nr
$\beta$ \|
0, ~~~0, ~~~~0, ~~~~1 \|
0, ~~~1, ~~~~1, ~~~~0 \|
0, ~~~1, ~~~~0, ~~~~1 \nr
$\gamma$ \|
1, ~~~1, ~~~~0, ~~~~0 \|\
1, ~~~1, ~~~~0, ~~~~0 \|
0, ~~~0, ~~~~0, ~~~~1  \endtable}
\hfill}
\smallskip
\parindent=0pt
\hangindent=39pt\hangafter=1
\baselineskip=18pt

{{\it Table 2.} A three generations ${SU(3)\times SU(2)\times U(1)^2}$
model. The choice of generalized GSO coefficients is:
${c\left(\matrix{b_1,b_3,\alpha, \beta,\gamma\cr
                                    \alpha\cr}\right)=
-c\left(\matrix{b_2\cr
                                    \alpha\cr}\right)=
c\left(\matrix{1,b_j,\gamma\cr
                                    \beta\cr}\right)=
-c\left(\matrix{\gamma\cr
                                    1,b_1,b_2\cr}\right)=
c\left(\matrix{\gamma\cr
                                    b_3\cr}\right)=
-1}$ (j=1,2,3),
with the others specified by modular invariance and space--time
supersymmetry.  The $16$ right--moving
internal fermionic states
$\{{\bar\psi}^{1,\cdots,5},{\bar\eta}^1,
{\bar\eta}^2,{\bar\eta}^3,{\bar\phi}^{1,\cdots,8}\}$,
correspond to the $16$ dimensional compactified  torus of the ten dimensional
heterotic string.
The 12 left--moving and 12 right--moving real internal fermionic states
correspond to the six left
and six right compactified dimensions in the bosonic language.}
\vskip 2.5cm

\vfill
\eject

\end
\bye

\input tables.tex
\nopagenumbers
\magnification=1000
\baselineskip=18pt
\hbox
{\hfill
{\begintable
\  \ \|\ ${\psi^\mu}$ \ \|\ $\{{\chi^{12};\chi^{34};\chi^{56}}\}$\ \|\
{${\bar\psi}^1$, ${\bar\psi}^2$, ${\bar\psi}^3$,
${\bar\psi}^4$, ${\bar\psi}^5$, ${\bar\eta}^1$,
${\bar\eta}^2$, ${\bar\eta}^3$} \ \|\
{${\bar\phi}^1$, ${\bar\phi}^2$, ${\bar\phi}^3$, ${\bar\phi}^4$,
${\bar\phi}^5$, ${\bar\phi}^6$, ${\bar\phi}^7$, ${\bar\phi}^8$} \crthick
$\alpha$
\|\ 0 \|
$\{0,~0,~0\}$ \|
1, ~~1, ~~1, ~~0, ~~0, ~~0 ,~~0, ~~0 \|
1, ~~1, ~~1, ~~1, ~~0, ~~0, ~~0, ~~0 \nr
$\beta$
\|\ 0 \| $\{0,~0,~0\}$ \|
1, ~~1, ~~1, ~~0, ~~0, ~~0, ~~0, ~~0 \|
1, ~~1, ~~1, ~~1, ~~0, ~~0, ~~0, ~~0 \nr
$\gamma$
\|\ 0 \|
$\{0,~0,~0\}$ \|
{}~~$1\over2$, ~~$1\over2$, ~~$1\over2$, ~~$1\over2$,
{}~~$1\over2$, ~~$1\over2$, ~~$1\over2$, ~~$1\over2$ \| $1\over2$, ~~0, ~~1,
{}~~1,
{}~~$1\over2$,
{}~~$1\over2$, ~~$1\over2$, ~~0 \endtable}
\hfill}
\smallskip
{\hfill
{\begintable
\  \ \|\
${y^3y^6}$,  ${y^4{\bar y}^4}$, ${y^5{\bar y}^5}$,
${{\bar y}^3{\bar y}^6}$
\ \|\ ${y^1\omega^5}$,  ${y^2{\bar y}^2}$,
${\omega^6{\bar\omega}^6}$,
${{\bar y}^1{\bar\omega}^5}$
\ \|\ ${\omega^2{\omega}^4}$,  ${\omega^1{\bar\omega}^1}$,
${\omega^3{\bar\omega}^3}$,  ${{\bar\omega}^2{\bar\omega}^4}$  \crthick
$\alpha$ \|
1, ~~~0, ~~~~0, ~~~~0 \|
0, ~~~0, ~~~~1, ~~~~1 \|
0, ~~~0, ~~~~1, ~~~~1 \nr
$\beta$ \|
0, ~~~0, ~~~~1, ~~~~1 \|
1, ~~~0, ~~~~0, ~~~~0 \|
0, ~~~1, ~~~~0, ~~~~1 \nr
$\gamma$ \|
0, ~~~1, ~~~~0, ~~~~1 \|\
0, ~~~1, ~~~~0, ~~~~1 \|
1, ~~~0, ~~~~0, ~~~~0  \endtable}
\hfill}
\smallskip
\parindent=0pt
\hangindent=39pt\hangafter=1
\baselineskip=18pt

{{\it Table 3.} A three generations ${SU(3)\times SU(2)\times U(1)^2}$
model. The choice of generalized GSO coefficients is:
${c\left(\matrix{b_j\cr
                                    \alpha,\beta,\gamma\cr}\right)=
-c\left(\matrix{\alpha\cr
                                    1\cr}\right)=
c\left(\matrix{\alpha\cr
                                    \beta\cr}\right)=
-c\left(\matrix{\beta\cr
                                    1\cr}\right)=
c\left(\matrix{\gamma\cr
                                    1,\alpha\cr}\right)=
-c\left(\matrix{\gamma\cr
                                    \beta\cr}\right)=
-1}$ (j=1,2,3),
with the others specified by modular invariance and space--time
supersymmetry.  The $16$ right--moving
internal fermionic states
$\{{\bar\psi}^{1,\cdots,5},{\bar\eta}^1,
{\bar\eta}^2,{\bar\eta}^3,{\bar\phi}^{1,\cdots,8}\}$,
correspond to the $16$ dimensional compactified  torus of the ten dimensional
heterotic string.
The 12 left--moving and 12 right--moving real internal fermionic states
correspond to the six left
and six right compactified dimensions in the bosonic language.}
\vskip 2.5cm

\vfill
\eject

\end
\bye

\input tables.tex
\nopagenumbers
\magnification=1000
\baselineskip=18pt
\hbox
{\hfill
{\begintable
\ &boundary conditions \  \|\ \ &~~$Z^{\pm}  \to$ ~~\  \crthick
 & (1) (1)            \  \|\ \ &$e^{i\pi}Z^{\pm}+2\pi+2\pi i$ \nr
 & (1) (2)            \  \|\ \ &$e^{i\pi}Z^{\mp}+2\pi+2\pi i$ \nr
 & (1) (3)            \  \|\ \ &$e^{i\pi}Z^{\pm}+2\pi$  \nr
 & (1) (4)            \  \|\ \ &$e^{i\pi}Z^{\mp}+2\pi+2\pi i$ \nr
 & (2) (1)            \  \|\ \ &$Z^{\mp}+2\pi+2\pi  i$ \nr
 & (2) (2)            \  \|\ \ &$Z^{\pm}+2\pi+2\pi i$ \nr
 & (2) (3)            \  \|\ \ &$Z^{\mp}+2\pi$ \nr
 & (3) (1)            \  \|\ \ &$e^{i\pi}Z^{\pm}+2\pi i$ \nr
 & (3) (2)            \  \|\ \ &$e^{i\pi}Z^{\mp}+2\pi i$ \nr
 & (3) (3)            \  \|\ \ &$e^{i\pi}Z^{\pm}$ \nr
 & (3) (4)            \  \|\ \ &$e^{i\pi}Z^{\mp}$ \nr
 & (4) (1)            \  \|\ \ &$Z^{\pm}+2\pi i$ \nr
 & (4) (2)            \  \|\ \ &$Z^{\pm}+2\pi i$ \nr
 & (4) (3)            \  \|\ \ &$Z^{\mp}$ \nr
 & (4) (4)            \  \|\ \ &$Z^{\pm}$
 \endtable}
\hfill}
\bigskip
\parindent=0pt
\hangindent=39pt\hangafter=1
{\it Table 4.} The transformations of the complex planes resulting from
combination of the world--sheet fermion boundary conditions which are
left--right symmetric. Here
$(1) \equiv y \to -y, \omega \to \omega$,
$(2) \equiv y \to -y, \omega \to -\omega$,
$(3) \equiv y \to y, \omega \to -\omega$,
$(4) \equiv y \to y, \omega \to \omega$.

\vfill
\eject

\end
\bye

\input tables.tex
\nopagenumbers
\magnification=1000
\baselineskip=18pt
\hbox
{\hfill
{\begintable
\ &boundary conditions \  \|\ \ &symmetry condition \  \|\ \ &remaining
moduli \crthick
 & (1) (3)            \  \|\ \ &$T=U$ \  \|\ \ &$H_1^1$ \nr
 & (3) (1)            \  \|\ \ &$T=-U$ \  \|\ \ &$H_2^1$ \nr
 & (1) (4)            \  \|\ \ &$T=-{\bar U}$ \  \|\ \ &$h_{21},h_{22}$ \nr
 & (4) (1)            \  \|\ \ &$T={\bar U}$ \  \|\ \ &$h_{11},h_{12}$ \nr
 & (2) (3)            \  \|\ \ &$T=-{\bar U}$ \  \|\ \ &$h_{21},h_{22}$ \nr
 & (3) (2)            \  \|\ \ &$T={\bar U}$ \  \|\ \ &$h_{11},h_{12}$ \nr
 & (2) (4)            \  \|\ \ &$T=U$ \  \|\ \ &$H_1^1$ \nr
 & (4) (2)            \  \|\ \ &$T=-U$ \  \|\ \ &$H_2^1$
 \endtable}
\hfill}
\bigskip
\parindent=0pt
\hangindent=39pt\hangafter=1
{\it Table 5.} The transformations of the complex planes resulting from
combination of the world--sheet fermion boundary conditions with one
of them left--right asymmetric. Here
$(1) \equiv X \to X$,
$(2) \equiv X \to -X$,
$(3) \equiv X \to X_L-X_R$,
$(4) \equiv X \to X_R-X_L$.
For clarity, the remaining moduli are given for the first complex plane.

\vfill
\eject

\end
\bye